\documentclass[reprint,prc,amsmath,amssymb,aps,showpacs,superscriptaddress]{revtex4-1}

\usepackage{natbib}
\usepackage{graphicx}% Include figure files
\usepackage{dcolumn}% Align table columns on decimal point
\usepackage{bm}% bold math
\usepackage{color}

\begin{document}

\preprint{APS/123-QED}

\title{Structure of $^{23}$Al from one-proton breakup reaction and astrophysical implications}

\author{A.~Banu}
\email[E-mail me at: ]{banula@jmu.edu}
\altaffiliation[Present address: ]{Department of Physics and Astronomy, James Madison University, Harrisonburg, VA 22807, USA}
\author{L.~Trache}
\affiliation{Cyclotron Institute, Texas A\&M University, College Station, TX 77843, USA}
\author{F.~Carstoiu}
\affiliation{National Institute of Physics and Nuclear Engineering ``Horia Hulubei'' (IFIN-HH), R-077125 Magurele-Bucharest, Romania}
\author{N. L.~Achouri}
\affiliation{LPC-ENSICAEN, IN2P3-CNRS et Universit\'e de Caen, 14050 Caen Cedex, France}
\author{A.~Bonaccorso}
\affiliation{Istituto Nazionale di Fisica Nucleare, Sez. di Pisa, I-56127 Pisa, Italy}
\author{W. N.~Catford}
\affiliation{Department of Physics, University of Surrey, Guildford GU2 5XH, UK}
\author{M.~Chartier}
\affiliation{Oliver Lodge Laboratory, University of Liverpool, Liverpool L69 7ZE, UK} 
\author{M.~Dimmock}
\affiliation{Oliver Lodge Laboratory, University of Liverpool, Liverpool L69 7ZE, UK} 
\author{B.~Fern\'andez-Dom\'inguez}
\altaffiliation[Present address: ]{LPC-ENSICAEN, IN2P3-CNRS et Universit\'e de Caen, 14050 Caen Cedex, France}
\affiliation{Oliver Lodge Laboratory, University of Liverpool, Liverpool L69 7ZE, UK}
\author{M.~Freer}
\affiliation{School of Physics and Astronomy, University of Birmingham, Birmingham B15 2TT, UK} 
\author{L.~Gaudefroy}
\altaffiliation[Present address: ]{CEA, DAM, DIF, F-91297 Arpajon Cedex, France}
\affiliation{Grand Acc\'el\'erateur d'Ions Lourds, BP 55027, 14076 Caen Cedex 5, France}  
\author{M.~Horoi}
\affiliation{Department of Physics, Central Michigan University, Mount Pleasant, Michigan 48859, USA}   
\author{M.~Labiche}
\affiliation{Departmet of Engineering and Science, University of the West of Scotland, Paisley, PA1 2BE, UK}
\affiliation{Nuclear Physics Group, STFC Daresbury Laboratory, Daresbury, Warrington, WA4 4AD, UK}
\author{B.~Laurent}
\altaffiliation[Present address: ]{CEA, DAM, DIF, F-91297 Arpajon Cedex, France}
\affiliation{LPC-ENSICAEN, IN2P3-CNRS et Universit\'e de Caen, 14050 Caen Cedex, France}
\author{R. C.~Lemmon}
\affiliation{Nuclear Physics Group, STFC Daresbury Laboratory, Daresbury, Warrington, WA4 4AD, UK}
\author{F.~Negoita}
\affiliation{National Institute of Physics and Nuclear Engineering ``Horia Hulubei'' (IFIN-HH), R-077125 Magurele-Bucharest, Romania}
\author{N. A.~Orr}
\affiliation{LPC-ENSICAEN, IN2P3-CNRS et Universit\'e de Caen, 14050 Caen Cedex, France}
\author{S.~Paschalis}
\altaffiliation[Present address: ]{Nuclear Science Division, LBNL, Berkeley, CA 94720, USA}
\affiliation{Oliver Lodge Laboratory, University of Liverpool, Liverpool L69 7ZE, UK} 
\author{N.~Patterson}
\affiliation{Department of Physics, University of Surrey, Guildford GU2 5XH, UK}
\author{E. S.~Paul}
\affiliation{Oliver Lodge Laboratory, University of Liverpool, Liverpool L69 7ZE, UK} 
\author{M.~Petri}
\altaffiliation[Present address: ]{Nuclear Science Division, LBNL, Berkeley, CA 94720, USA}
\affiliation{Oliver Lodge Laboratory, University of Liverpool, Liverpool L69 7ZE, UK} 
\author{B.~Pietras}
\affiliation{Oliver Lodge Laboratory, University of Liverpool, Liverpool L69 7ZE, UK}  
\author{B. T.~Roeder}
\altaffiliation[Present address: ]{Cyclotron Institute, Texas A\&M University, College Station, TX 77843, USA}
\affiliation{LPC-ENSICAEN, IN2P3-CNRS et Universit\'e de Caen, 14050 Caen Cedex, France}
\author{F.~Rotaru}
\affiliation{National Institute of Physics and Nuclear Engineering ``Horia Hulubei'' (IFIN-HH), R-077125 Magurele-Bucharest, Romania}
\author{P.~Roussel-Chomaz}
\altaffiliation[Present address: ]{CEA-Saclay, DSM/IRFU/SPhN, F-91191, Gif-sur-Yvette, France}
\affiliation{Grand Acc\'el\'erateur d'Ions Lourds, BP 55027, 14076 Caen Cedex 5, France}   
\author{E.~Simmons}
\affiliation{Cyclotron Institute, Texas A\&M University, College Station, TX 77843, USA}
\author{J. S.~Thomas}
\altaffiliation[Present address: ]{School of Physics and Astronomy, The University of Manchester, Manchester M13 9PL, UK}
\affiliation{Department of Physics, University of Surrey, Guildford GU2 5XH, UK}
\author{R. E.~Tribble}
\affiliation{Cyclotron Institute, Texas A\&M University, College Station, TX 77843, USA}


\begin{abstract}
The ground state of the proton-rich nucleus $^{23}$Al has been studied by one-proton removal on a carbon target at about 50 MeV/nucleon using the EXOGAM + SPEG experimental setup at GANIL. Longitudinal momentum distributions of the $^{22}$Mg breakup fragments, inclusive and in coincidence with gamma rays de-exciting the residues, were measured. The ground-state structure of $^{23}$Al is found to be a configuration mixing of a \textit{d}-orbital valence proton coupled to four core states - 0$^{+}_{gs}$, 2$^{+}_{1}$, 4$^{+}_{1}$, 4$^{+}_{2}$. We confirm the ground state spin and parity of $^{23}$Al as $J^{\pi} = 5/2^{+}$.  The measured exclusive momentum distributions are compared with extended Glauber model calculations to extract spectroscopic factors and asymptotic normalization coefficients (ANCs). The spectroscopic factors are presented in comparison with those obtained from large-scale shell model calculations. We determined the asymptotic normalization coefficient of the nuclear system $^{23}$Al$_{gs}$ $\rightarrow$ $^{22}$Mg(0$^{+}$) + p to be $C^{2}_{d_{5/2}}$($^{23}Al_{gs}$) = (3.90 $\pm$ 0.44) $\times$ 10$^{3}$ fm$^{-1}$, and used it to infer the stellar reaction rate of the direct radiative proton capture $^{22}$Mg(p,$\gamma$)$^{23}$Al. Astrophysical implications related to $^{22}$Na nucleosynthesis in ONe novae and the use of one-nucleon breakup at intermediate energies as an indirect method in nuclear astrophysics are discussed.
\end{abstract}

% PACS, the Physics and Astronomy Classification Scheme

\maketitle

\section{\label{sec:level1}Introduction}

 The proton-rich nucleus $^{23}$Al near the dripline, which was first identified in 1969 \cite{cern69}, has been known early on \cite{goug72} as a $\beta$-delayed proton emitter. However, detailed or precise information on its structure and decay scheme is scarce. Only a few years back, basic information such as the spin and parity of its ground state was still missing, and its mass was uncertain even in the most recent compilations. It became more accessible for studies in the last few years owing to its availability as projectile or source due to better separation techniques, and a number of publications ensued \cite{wies88,cagg01,xcai02,zhan02,gomi05,ozaw06,iaco06,fang07,jjhe07,gade08,abdu10}. $^{23}$Al has been found an appealing nucleus raising questions related to both nuclear structure and nuclear astrophysics, as highlighted in the following paragraphs. \newline
\indent Recent measurements of reaction cross sections for $N$ = 10 isotones \cite{xcai02,zhan02} and $Z$ = 13 isotopes \cite{xcai02,fang07} on a carbon target showed a slight enhancement (of about 10$\%$) for $^{23}$Al. This has led to the interpretation of $^{23}$Al as a proton halo nucleus, and the authors went to a great length to demonstrate it \cite{zhan02a}. For an $sd$-shell nucleus, this feature could only be explained by assuming a level inversion between the proton $2s_{1/2}$ and $1d_{5/2}$ orbitals. Before these results were published, the assumption was that the spins and parities in $^{23}$Al were the same as in the mirror nucleus $^{23}$Ne, which has a ground state spin and parity of $J^{\pi} = 5/2^{+}$ and 1/2$^{+}$ for the first excited state. The level inversion in $^{23}$Al would imply that its ground state is $J^{\pi} = 1/2^{+}$. Thus, the exotic proton halo structure in $^{23}$Al could be explained, but the mirror symmetry would be broken for bound ground states of mirror pairs. This would be the first case of such a signature with significant implications not only for nuclear physics but also for nuclear astrophysics \cite{trac06}.\newline
\indent Space-based $\gamma$-ray telescopes have shown the ability to detect gamma rays of cosmic origin providing us with a direct evidence that nucleosynthesis is an on-going process in our galaxy. Gamma rays emitted by long-lived isotopes, such as $^{26}$Al (T$_{1/2}$ = 0.7 $\times 10^{6}$ yr) or $^{60}$Fe (T$_{1/2}$ = 1.5 $\times 10^{6}$ yr), have been observed. Among the proposed $\gamma$-ray emitters of cosmic origin is also the shorter lived $^{22}$Na (T$_{1/2}$ = 2.6 yr) \cite{clay74}, predicted to be synthesized in explosive ONe novae at temperatures between 0.2 and 0.4 billion Kelvin through the reaction path $^{20}$Ne(p,$\gamma$)$^{21}$Na(p,$\gamma$)$^{22}$Mg($\beta^{+}\nu$)$^{22}$Na. However, the $\gamma$-ray line of 1.275 MeV following the $\beta^{+}$-decay of $^{22}$Na from novae has not been observed yet by state-of-the-art space-based  telescopes such as COMPTEL \cite{iyun95} or INTEGRAL \cite{diehl03}. Hence, novae models and/or nuclear data could be questioned. It has been proposed that $^{22}$Na itself and/or its precursor $^{22}$Mg could be depleted by $^{22}$Na(p,$\gamma$)$^{23}$Mg and $^{22}$Mg(p,$\gamma$)$^{23}$Al radiative capture reactions. We treat here the latter radiative capture. The $^{22}$Mg(p,$\gamma$)$^{23}$Al reaction is dominated by non-resonant capture to the ground state of $^{23}$Al and by resonant capture to its first excited state \cite{wies88, cagg01, gomi05}.\newline 
\indent It has been demonstrated that the momentum distributions of the core fragments measured in one-nucleon breakup reactions (we favor the term $breakup$ over $knockout$ used by other groups \cite{hans01}) are powerful spectroscopic tools to determine the single-particle structure of the nuclei far from stability. The shapes (widths) of these momentum distributions provide information on the orbital angular momentum \textit{l} of the removed nucleon \cite{hans01,sauv04}, whereas the nuclear breakup cross section determines the asymptotic normalization coefficient. This ANC is used to calculate the direct (non-resonant) component of the astrophysical $S$ factor of the radiative capture reaction \cite{trac01}. To accomplish this for the $^{22}$Mg(p,$\gamma$)$^{23}$Al direct capture, the configuration mixing in the $^{23}$Al ground state has to be determined, and careful cross section calculations have to be performed.\newline
\indent We have proposed to determine the spin of $^{23}$Al, its structure and the ANC for $^{23}$Al$_{gs}$ $\rightarrow$ $^{22}$Mg(0$^{+}$) + p by measuring inclusive and exclusive momentum distributions using nuclear breakup at intermediate energies. Meanwhile, a recent experiment at RIKEN has found that the magnetic moment of $^{23}$Al$_{gs}$ \cite{ozaw06} is only compatible with a spin 5/2. In parallel, it has been found unambiguously from the $\beta^{+}$-decay of $^{23}$Al \cite{iaco06} that its ground-state spin-parity is $J^{\pi}=5/2^{+}$, the same as for its mirror nucleus $^{23}$Ne.\newline
\indent In this paper, we describe the determination, with an independent experimental method, of the structure of $^{23}$Al. Thus, in addition to confirming the ground-state spin and parity of $^{23}$Al, we provide for the first time the information on the configurations that make up the ground state of $^{23}$Al. The configuration mixing is obtained by the use of coincidences with $\gamma$-rays from the $^{22}$Mg core residues, left excited after the one-proton breakup of $^{23}$Al. The comparison between experimental momentum distributions and calculations enables us to extract the corresponding spectroscopic factors and the ANC for $^{23}$Al$_{gs}$ $\rightarrow$ $^{22}$Mg(0$^{+}$) + p. The experimental spectroscopic factors are compared with those obtained from large-scale shell model calculations made with modern effective interations. Using the ANC, the astrophysical $S$ factor for the $^{22}$Mg(p,$\gamma$)$^{23}$Al reaction is evaluated. It is believed that novae could become the first type of explosive process for which all the nuclear input to the nucleosynthesis calculations is based on experimental data \cite{iliad02}, and this work is a step in that direction with a new method.

\section{\label{sec:level2}Experiment}
The experiment was performed at the GANIL coupled cyclotron facility. A cocktail of secondary beams was produced via the fragmentation of an intense ($\sim$ 2 $\mu$A) 95 MeV/nucleon $^{32}$S$^{16+}$ primary beam on a thick carbon target. The secondary ion beams were collected using the SISSI device \cite{anne97} coupled to a beam analysis spectrometer tuned at $B\rho$ = 1.954 Tm, and operated with a Beryllium achromatic degrader. Fourteen ion species - $^{13}$B, $^{14}$C, $^{15,16}$N, $^{16,17}$O, $^{18,19}$F, $^{19,20}$Ne, $^{21}$Na, $^{22}$Mg, $^{23}$Al, and $^{24}$Si - with energies between 24 and 60 MeV/nucleon and intensities ranging from ~30 and 7000 pps were obtained. We had about 300 pps of $^{23}$Al at 57 MeV/nucleon. A secondary reaction target of carbon, 175-mg/cm$^{2}$ thick, was used. To measure the breakup fragment momentum distributions, the SPEG spectrometer \cite{bian89} was employed and operated at 0$^{\circ}$ in an achromatic mode on target, whereby an intrinsic resolution of $\delta$p/p $\sim$ 5$\times$10$^{-4}$ (FWHM) was achieved. The final momentum resolution, including target effects, was $\delta$p/p $\sim$ 5$\times$10$^{-3}$ (FWHM). With a large angular acceptance of 4$^{\circ}$ in both horizontal and vertical planes, the overall momentum acceptance of the spectrometer was 7$\%$. This permitted the momentum distributions of the fragments resulting from one-proton breakup of all nuclei of interest to be measured in a single setting. SPEG was tuned for the magnetic rigidity of $^{22}$Mg residues ($B_{\rho}^{SPEG}$ = 1.756 Tm).\newline
\indent Ion identification at the focal plane of SPEG was achieved using the energy loss from an ionization gas chamber and the time-of-flight between a thick plastic stopping detector and the cyclotron radio frequency. Two large-area drift chambers straddling the focal plane of SPEG allowed the focal plane position spectra to be reconstructed. The longitudinal momentum of each particle was derived from the reconstructed focal plane position. The momentum of the core fragment relative to the incident projectile in the laboratory frame was transformed into that in the projectile rest frame using Lorentz transformation. To compare the measured distributions with the theoretical ones, all broadening effects inherent in the measurements have been taken into account through Monte Carlo simulations. These effects include the energy spread in the beam, the differential energy losses of the projectile and the fragment in the target, the energy and angular straggling in the target, and the detector and spectrometer resolutions.\newline
\indent The reaction target was surrounded by 8 EXOGAM \cite{shep99} Germanium clover detectors set up in a new configuration, for the first time in association with SPEG. The absolute efficiencies of the EXOGAM detectors were determined using calibrated $\gamma$-ray sources ($^{152}$Eu, $^{56,60}$Co, $^{137}$Cs). The array was used in a configuration with four detectors at 45$^{\circ}$ (forward angles) and the other four at 135$^{\circ}$ (backward angles) with respect to the beam axis, at 215 and 134 mm from target, respectively.  This configuration resulted in an efficiency of $\sim$ 3$\%$ at 1.33 MeV \cite{piet09}. To determine the absolute efficiencies of the EXOGAM array for the detection of gamma rays de-exciting the breakup fragments, Lorentz tranformation for in-flight emitted gamma rays was applied. We assumed an isotropic $\gamma$-ray emission in the projectile reference frame. One notes that at relativistic energies, the so-called Lorentz boost plays a major role in increasing the detection efficiency of $\gamma$-rays emitted at forward angles.\newline
Each of the EXOGAM clovers is 16-fold segmented and allowed for an event-by-event addback and Doppler reconstruction of the gamma rays emitted in-flight. The emission angle employed for the Doppler correction was determined from the location of the segment with the largest energy deposition.\newline
\indent The intensities of the secondary beams were derived from several empty-target normalization runs, with respect to the known primary beam intensity. The final cross sections were determined using an average of these normalization runs. We estimated that the normalization uncertainty is 11\%.

\section{\label{sec:level3}Results}
We have measured the inclusive and exclusive longitudinal momentum distributions of the $^{22}$Mg breakup fragments and the corresponding differential and integral breakup cross sections. By detecting the $\gamma$-ray decays of excited states in $^{22}$Mg residues, we were able to disentangle for the first time the configuration mixing structure of the $^{23}$Al ground state. With a small proton-separation energy of $S_{p}$ = 141.11(43) keV \cite{saas09} (the most accurate value to date) compared with that of $^{22}$Mg (5502 (2) keV \cite{audi03}), the low-lying nuclear structure of $^{23}$Al can be assumed to be that of a core nucleus plus a valence proton ($^{22}$Mg + p).\newline
\indent The measured inclusive momentum distribution is compared in Fig. 1 with extended Glauber-type calculations, which are explained in detail in Ref. \cite{sauv04} and briefly discussed later in this paper. For the calculations, everywhere we used $^{23}$Al mid-target energy of 50 MeV/nucleon. The single-particle wave functions are calculated in a Woods-Saxon proton binding potential with a set of radius and diffuseness parameters $r_{0}$ = 1.18 fm and $a$ = 0.60 fm. With this Woods-Saxon potential, the theoretical inclusive momentum distribution (full curve) is calculated in the $J^{\pi} = 5/2^{+}$ hypothesis based on the $1d_{5/2}$ orbital, using the core configurations and spectroscopic factors predicted by large-scale shell model calculations (see Table 1, columns 1 and 7) with the USDB effective interaction \cite{brow06}. \textit {We underline the calculation is not a fit.} It does not only reproduce the shape and width, but also the absolute value of the cross section. For the integral theoretical cross section, we obtained a value of $\sigma_{inc}^{th}$ = 77.7 mb, whereas the corresponding experimental integral cross section, corrected for the missing counts on the leftmost low momentum tail of the longitudinal momentum distribution, amounts to $\sigma_{inc}$ = 78.3(4) mb.\newline
\indent It is clear that the width (FWHM) of 180 MeV/c of the measured core momentum distribution agrees with that calculated with the $1d_{5/2}$ orbital, but not for the $|^{22}Mg(0^{+}_{gs})\otimes \pi2s_{1/2}\rangle$ case (about 60 MeV/c; dashed curve, \textit{arbitrary normalization}). From here we confirm that the spin-parity for the ground state of $^{23}$Al is $J^{\pi}$=5/2$^{+}$, as expected, and the same as for its mirror nucleus $^{23}$Ne.\newline
%--- Figure 1 ----
\begin{figure}[h]
\includegraphics[scale=0.30,keepaspectratio=true]{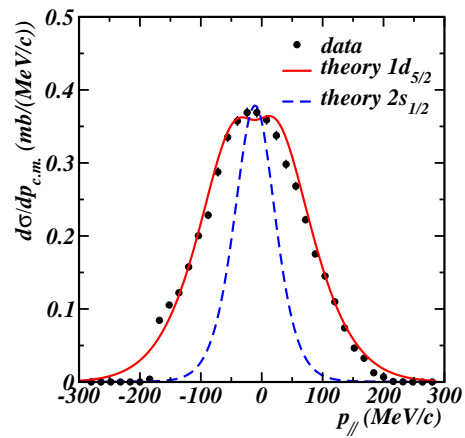}
\caption[Fig. 1]{(Color online) Experimental inclusive momentum distribution of $^{22}$Mg cores (points), in the center-of-mass frame, compared with a theoretical distribution calculated for the $2s_{1/2}$ single-particle orbital (dashed curve, arbitrary units) and with a calculated inclusive momentum distribution using the $1d_{5/2}$ orbital and the theoretical spectroscopic factors (full curve, absolute normalization) obtained from large-scale shell model calculations (see text).}
\end{figure}
\indent The Doppler corrected $\gamma$-ray energy spectrum in Fig. 2 is obtained from data taken in coincidence with $^{22}$Mg breakup fragments. We have identified three $\gamma$-ray lines - 1247 keV, 2061 keV, and 1985 keV \cite{NNDC}- corresponding, respectively, to the transitions 2$^{+}_{1}$ $\rightarrow$ 0$^{+}_{gs}$, 4$^{+}_{1}$ $\rightarrow$ 2$^{+}_{1}$, and the less expected 4$^{+}_{2}$ $\rightarrow$ 4$^{+}_{1}$ (see inset).
%--- Figure 2 ----
\begin{figure}
\begin{center}
\includegraphics[scale=0.45,keepaspectratio=true]{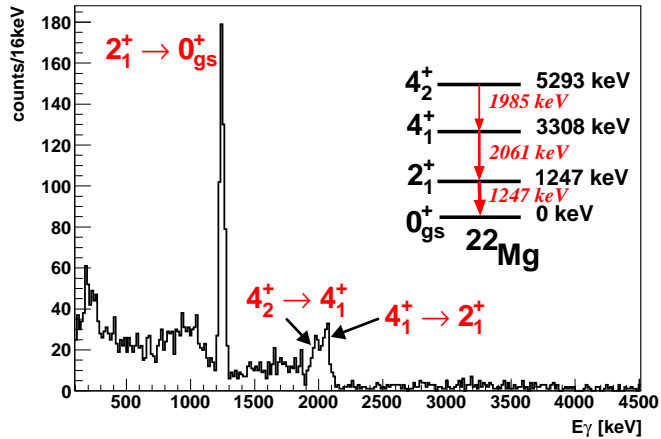}
\caption[Fig. 2]{(Color online) Doppler-corrected $\gamma$-ray spectrum in coincidence with identified $^{22}$Mg residues in SPEG. The inset shows the levels and transitions in $^{22}$Mg core fragments.}
\end{center}
\end{figure} 
This leads to a configuration mixing in the ground state of $^{23}$Al of the type:
$|^{23}Al_{gs}(5/2^{+})\rangle$ = $A_{0}|^{22}Mg(0^{+}_{gs})\otimes \pi1d_{5/2}\rangle$ + $A_{1}|^{22}Mg(2^{+}_{1})\otimes \pi1d_{5/2}\rangle$ + $A_{2}|^{22}Mg(4^{+}_{1})\otimes \pi1d_{5/2}\rangle$ + $A_{3}|^{22}Mg(4^{+}_{2})\otimes \pi1d_{5/2}\rangle$, where $A_{i}$ (i=0-3) represent the spectroscopic amplitudes of each of the four configurations. The last three components can also couple the $1d_{3/2}$ proton orbital to the $^{22}$Mg core, which can not be excluded based on the measured momentum distributions, sensitive only to the orbital angular momentum. However, the shell-model calculations showed that the associated spectroscopic factors are a factor of at least 50 smaller for these components. Therefore, they are neglected here (Fig. 1 validates this option).\newline
\indent The exclusive momentum distributions are shown in Fig. 3 compared with theoretical calculations. In the data analysis, background subtraction was applied in each case, and the feeding contributions were considered for the $4^{+}_{1} \rightarrow 2^{+}_{1}$ and $2^{+}_{1} \rightarrow 0^{+}_{gs}$ transitions in the $\gamma$-ray cascade of de-excitation to the ground state in $^{22}$Mg. We note in Table 1 (column 4) the large momentum widths, about 200 MeV/c, characteristic to the $1d_{5/2}$ valence proton orbital. The increase in width with core excitation energy of the exclusive momentum distributions is expected as the excitation of the core increases the effective binding energy of the valence proton.\newline
The momentum distribution corresponding to $^{22}$Mg ground state (top panel in Fig. 3) was derived by subtraction of the measured exclusive momentum distributions from the measured inclusive momentum distribution. We stress here that an important factor in the procedure and in the uncertainty estimations was a good knowledge of the absolute efficiencies for the in-flight detection of the three gamma-ray lines.\newline
%--- Figure 3 ----
\begin{figure}
\includegraphics[scale=0.5,keepaspectratio=true]{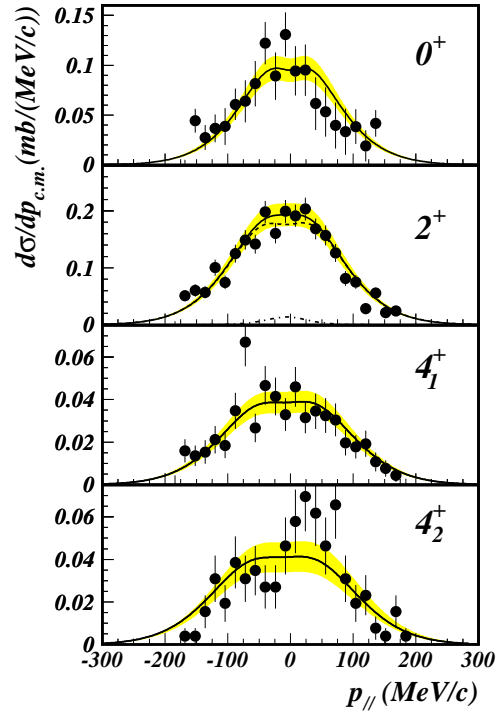}
\caption[Fig. 3]{(Color online) Experimental exclusive momentum distributions determined in the center-of-mass frame for $^{22}$Mg residues corresponding to $^{23}$Al ground state configuration mixing. Comparison with calculations using spectroscopic factors from fit (see text). In the second panel from the top the full (dashed) curve is associated with calculations that include (exclude) the contribution of the 
$|^{22}Mg(2^{+})\otimes \pi2s_{1/2}\rangle$. The dot-dashed curve is the calculated momentum distribution of a pure $s$ wave. Shaded areas correspond to $1\sigma$ deviation in the spectroscopic amplitudes. The uncertainties contain the statistical errors and those from the $\gamma$-ray efficiencies.}
\end{figure} 
\indent In the second panel from the top in Fig. 3, the measured and fitted momentum distributions of the $|2^{+}\otimes \pi1d_{5/2}\rangle$ configuration agree with each other, and no significant $|2^{+}\otimes \pi2s_{1/2}\rangle$ component is possible to explain the proposed halo structure. The upper limit for the spectroscopic factor coresponding to the $|2^{+}\otimes \pi2s_{1/2}\rangle$ configuration, deduced from our data, is 0.09. This is a marginal contribution to the wave function of $^{23}$Al, which clearly can not be a halo nucleus.\newline
\indent Figure 4 illustrates the contribution of each of the configurations identified in the ground state of $^{23}$Al to the inclusive longitudinal momentum distribution. The curves are calculated with the Glauber-type reaction model described below, normalized with the fitted spectroscopic factors  (Table 1, column 6), and the points are the experimental data. Although the $|2^{+}\otimes \pi2s_{1/2}\rangle$ configuration has a minor contribution in the configuration mixing, it provides a better fit to the data when included in the full calculation.\newline
%--- Figure 4 ----
\begin{figure}
\includegraphics[scale=0.5,keepaspectratio=true]{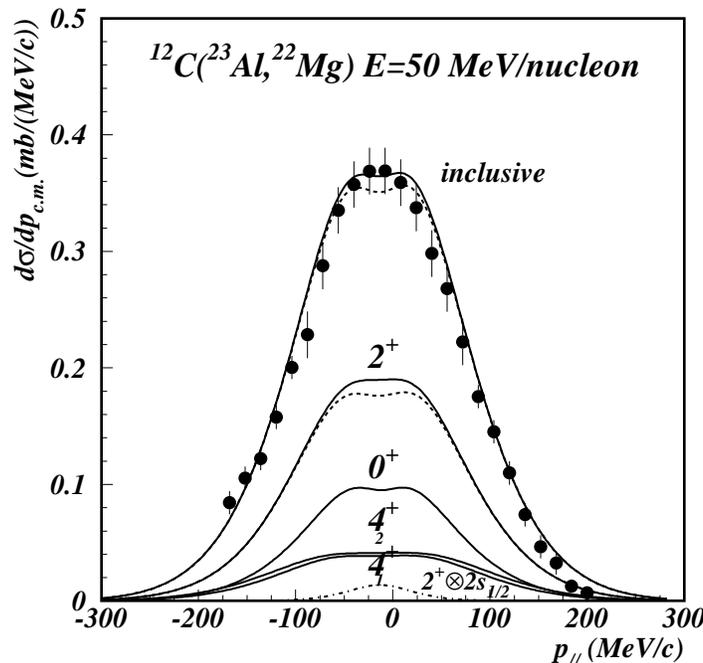}
\caption[Fig. 4]{Experimental inclusive momentum distribution (points) in the center-of-mass reference frame compared with the calculated one using the fitted spectroscopic factors (full curve). The lower curves present the contributions of each of the configurations identified, labelled by the core states. The full (dashed) curve is associated with the full calculation that includes (excludes) the contribution of the $|^{22}Mg(2^{+})\otimes \pi2s_{1/2}\rangle$. Same for the curve labelled 2$^{+}$.}
\end{figure}
\setlength{\tabcolsep}{0.4cm}
\begin{table*}[!]
\begin{center}
\caption{\label{table1} Cross sections, widths (FWHM) of momentum distributions, asymptotic normalization coefficients and spectroscopic factors for one-proton removal from $^{23}$Al. The experimental spectroscopic factors, $S_{exp}$, and the theoretical cross sections are obtained from the bootstrap procedure (see text). The theoretical spectroscopic factors, $S_{th}$, are from large-scale shell model calculations using the USDB effective interaction \cite{brow06} with a center-of-mass correction $(\frac{A}{A-1})^2$ \cite{diep74} applied. The uncertainties contain only statistical and calculation contributions, but not the overall 11\% uncertainty in the cross section normalization.}
\begin{tabular}{cccccccc} \hline
configuration & $E_{\gamma} $ & $\sigma_{-1p}^{exp}$  & FWHM  & $C^{2}$ & $S_{exp}$ & $S_{th}$ & $\sigma_{-1p}^{fit}$ \\
              &    [keV]           &       [mb]                &     [MeV/c]  &    [fm$^{-1}$] &            &               &         [mb]             \\
\hline
$^{22}Mg(0^{+}_{gs})\otimes \pi_{1d_{5/2}}$ & 0    & 18.5 $\pm$ 1.2 & 160 $\pm$ 5 & 3896 $\pm$ 113 & 0.45 $\pm$ 0.07 & 0.36  & 19.1 $\pm$ 2.0\\
$^{22}Mg(2^{+}_{1})\otimes \pi_{1d_{5/2}}$  & 1247 & 39.3 $\pm$ 1.2 & 180 $\pm$ 11 & 10.4 $\pm$ 1.0 & 1.15 $\pm$ 0.18 & 0.92  & 39.0 $\pm$ 2.2\\
$^{22}Mg(2^{+}_{1})\otimes \pi_{2s_{1/2}}$  & 1247 &              &     &               & $<$ 0.09      & 0.003 &              \\
$^{22}Mg(4^{+}_{1})\otimes \pi_{1d_{5/2}}$  & 2061 & 9.5 $\pm$ 0.9  & 200 $\pm$ 8 & 5.4 $\pm$ 0.8 & 0.34 $\pm$ 0.06 & 0.27  &  9.3 $\pm$ 0.9\\
$^{22}Mg(4^{+}_{2})\otimes \pi_{1d_{5/2}}$  & 1985 & 10.9 $\pm$ 0.9 & 210 $\pm$ 7 & 13.8 $\pm$ 2.2 & 0.50 $\pm$ 0.09 & 1.43  & 10.4 $\pm$ 1.6\\
\textbf{inclusive}                                   &      & \textbf{78.3 $\pm$ 0.4} & \textbf{180 $\pm$ 9} &  &                &      & \textbf{77.8 $\pm$ 0.7}\\
\hline
\end{tabular}
\end{center}
\end{table*} 
\indent In the extended Glauber model \cite{sauv04}, applied here for describing the nuclear breakup reactions, the cross sections are calculated as an incoherent sum of single-particle components,
\begin{equation}
\sigma_{-1p}=\sum S(c;nlj)\sigma_{sp}(nlj),
\label{eq1}
\end{equation}
where the sum extends over the single particle quantum numbers $nlj$ of the orbitals contributing for a given core state $c$, $S$ are the spectroscopic factors and $\sigma_{sp}$ are the single-particle removal cross sections. A similar relation holds for the momentum distributions. If the breakup reaction is peripheral, Eq. (1) can be re-written in terms of asymptotic normalization coefficients taking into account the relationship 
\begin{equation}
S(c;nlj)= C^{2}(c;nlj)/b^{2}_{sp}(nlj),
\end{equation}
where $C(c;nlj)$ and $b_{sp}$ are the ANC of the system $^{23}$Al$ \rightarrow$ $^{22}$Mg + p, and the single-particle ANC, respectively.\newline
\indent To extract the ANCs and the spectroscopic factors, a robust bootstrap procedure \cite{boot00} was applied. A standard objective $\chi^2$ function was defined for each observable using the experimental uncertainties. The total $\chi^2_{tot}$ is minimized searching for the spectroscopic factors used in Eq. (\ref{eq1}). We varied the optical potentials and the geometry of the proton binding potential used in the single-particle cross section calculations, as described below.\newline
\indent Coulomb dissociation is calculated in first-order perturbation theory, including final-state interactions. The optical potentials for the core-target and the proton-target systems are generated by folding the density- and energy-dependent microscopic interaction of Jeukenne, Lejeune and Mahaux (JLM) \cite{jeuk77}. The single-particle densities for the core and target used here were obtained in a standard spherical HF + BCS calculation using the density functional of Beiner and Lombard \cite{bein74}. The core rms charge radius obtained in this calculation for $^{22}$Mg core is $<r^2_{ch}>^{1/2}$=3.05 fm, which compares well with the experimental value for $^{24}$Mg (3.075$\pm$0.015 fm) \cite{ange98}. The calculated ${\it rms}$ charge radius of the $^{12}$C target  is almost identical with the experimental value (2.470$\pm$0.002 fm) \cite{ange98}. Renormalization of the real and imaginary optical potentials were choosen randomly distributed within $3\sigma$ deviation of the values found in Ref. \cite{opti00}. These renormalizations were tested in detail against $^{22}$Ne+$^{12,13}$C elastic scattering at 12 MeV/nucleon \cite{abdu10}. We assumed that the remaining energy dependence of the optical potentials is well acounted for by the intrinsic energy dependence of the JLM effective interaction. The S-matrix elements in impact parameter representation, defining the transition operators for stripping and diffraction, were calculated in the eikonal approximation including noneikonal corrections up to second order \cite{cars93}. The $1d_{5/2}$ wave functions for the valence proton were generated in a spherical Woods-Saxon (WS) potential with a radius and diffuseness randomly distributed in the ranges $R$ = 3.0-3.6 fm and $a$ = 0.50-0.70 fm by adjusting the depth of the potential to reproduce the effective separation energy $E_{eff}=S_{p}+E_x(I_c)$, where $S_{p}$ is the experimental proton separation energy for the ground state and $E_x$ is the experimental core excitation energy. The spin-orbit component was 
taken in the Thomas form with a standard strength, while the Coulomb component was generated by a uniform charge distribution with a
radius equal to the nuclear value.\newline
\indent We extract the ANCs and the spectroscopic factors, and evaluate the uncertainties due to statistics and calculations. Their values are listed in Table 1 (columns 5 and 6) and used to calculate the theoretical curves displayed in Fig. 3. The ANCs and the spectroscopic factors are weighted averages, and the uncertainties reflect their dependence on the binding potentials used. While for the ANC of the ground state, the uncertainty is less than 3\%, the uncertainty of the corresponding spectroscopic factor is 16\%. The ANC for the $|0^{+}_{gs}\otimes \pi1d_{5/2}\rangle$ component is found $C^{2}_{d_{5/2}}(^{23}Al_{gs})_{stat}$ = $(3.90\pm 0.11)\times$10$^{3}$ fm$^{-1}$. Adding the 11\% uncertainty in the overall normalization of the experimental cross sections, we obtain $C^{2}_{d_{5/2}}(^{23}Al_{gs})$ = $(3.90\pm 0.44)\times$10$^{3}$ fm$^{-1}$, which is the ANC for the system $^{23}$Al$_{gs}$ $\rightarrow$ $^{22}$Mg(0$^{+}$) + p of interest here.\newline
\indent The  experimental spectroscopic factors are in reasonably good agreement with those obtained from large-scale shell model calculations based on the USDB effective interaction \cite{brow06}. The discrepancies are within the limits found by recent surveys of the spectroscopic factors derived from light ion transfer reactions for $sd$- and $pf$-shell nuclei \cite{tsan09}. The sum of the spectroscopic factors listed in Table 1 (column 7) for $^{23}$Al exhausts 70\% of the 1$d_{5/2}$ proton occupation number of 4.3, predicted by the shell model. As in previous cases, we have determined that the ANC is less dependent on the parameters (i.e. the geometry) of the proton binding potential used in the calculation of the breakup cross sections than the extracted spectroscopic factors.\newline 
\indent In Ref. \cite{abdu10}, the authors studied the neutron transfer reaction $^{13}$C($^{22}$Ne,$^{23}$Ne)$^{12}$C, and determined the ANC for the $\nu 1d_{5/2}$ component in the system $^{23}$Ne$\rightarrow ^{22}$Ne+n. Based on the assumption that neutron and proton spectroscopic factors are equal in mirror nuclei, the value for the ANC of the mirror system $^{23}$Al$\rightarrow ^{22}$Mg+p was C$^{2}_{d5/2}(^{23}Al_{gs}) = 4.63(77)\times 10^{3}$ fm$^{-1}$, which is in agreement, within the uncertainties, with the value obtained here directly. This agreement supports the assumption that spectroscopic factors are equal in mirror nuclei, even when one of them is close to the dripline.

\section{\label{sec:level4}Astrophysical implications}
Of the four configurations contributing significantly to the structure of the $^{23}$Al ground state, the one relevant for the $^{22}$Mg(p,$\gamma$)$^{23}$Al reaction in stars is the component based on the $^{22}$Mg ground state. We used the corresponding ANC to evaluate the non-resonant component of the astrophysical $S$ factor, $S_{dir}(E)=0.73+0.17\cdot E+0.43\cdot E^{2}-0.21\cdot E^{3} keV\cdot b$, for energies $E$ = 0-1 MeV. From it, we evaluated the contribution to the stellar reaction rate for temperatures $T$ = 0-1 GK. The resonant contribution due to the capture through the first excited state was calculated with $E_{res}=528(19)-141.11(43)=387(19)$ keV (using the excitation energy from Ref. \cite{cagg01} and the new value for the proton binding energy \cite{saas09}) and the resonant strength $\omega\gamma=1/3\cdot 7.2\times10^{-7}$ eV from the Coulomb dissociation of $^{23}$Al \cite{gomi05}. 
We find S$(0)=0.73\pm0.10$ keV$\cdot$b. The values found are in agreement with those determined in the latest analysis of the reaction rate for $^{22}$Mg(p,$\gamma$)$^{23}$Al presented in Ref. \cite{abdu10}, and not far from those evaluated earlier by Wiescher et al. \cite{wies88} and Caggiano et al.  \cite{cagg01}. It follows that proton capture on $^{22}$Mg is not capable of explaining the depletion of $^{22}$Na produced in novae. The answer should be sought in other reactions, such as $^{22}$Na(p,$\gamma$)$^{23}$Mg (the capture on $^{22}$Na itself), which is larger and dominated by a number of still poorly known resonances. Moreover, its low proton separation energy means that $^{23}$Al is very easily destroyed by photo-desintegration and the two mechanisms equilibrate quickly. The slightly larger proton separation energy ($S_p=141$ keV), recently found, leads to about a factor two increase in the equilibrium  $^{23}$Al density at T$_9=0.3$ in comparison with the previous estimates (for $S_p=123$ keV), and it may help in the sequential two-proton capture on $^{22}$Mg in higher temperature and density environments, such as X-ray bursters. However, in novae the contribution of the $^{22}$Mg(p,$\gamma$)$^{23}$Al capture remains marginal.

\section{\label{sec:level5}Conclusions}
In this work, we have studied the ground state structure of $^{23}$Al using one-proton breakup reaction at intermediate energies and extracted the asymptotic normalization coefficient of the nuclear system $^{23}$Al$_{gs}$ $\rightarrow$ $^{22}$Mg(0$^{+}$) + p. For the first time, the configuration mixing in a complex case was determined from one-nucleon breakup using high-resolution segmented Germanium detectors. We extracted the components of the $^{23}$Al ground state wave function from measured inclusive and exclusive momentum distributions of the breakup fragments, and showed that the ground state of $^{23}$Al is dominated by configurations consisting of a valence 1$d_{5/2}$ proton coupled to low-lying states in $^{22}$Mg. We found no ground that $^{23}$Al is a halo nucleus as claimed previously, and confirmed the ground state spin and parity of $^{23}$Al as $J^{\pi} = 5/2^{+}$. Experimentally extracted spectrosopic factors for each of the measured core configurations compare reasonably with those from shell-model calculations. The value of the asymptotic normalization coefficient of $^{23}$Al$_{gs}$ $\rightarrow$ $^{22}$Mg(0$^{+}$) + p extracted here directly from one-proton breakup reaction is used to evaluate the astrophysical $S$ factor and the reaction rate for $^{22}$Mg(p,$\gamma$)$^{23}$Al. We conclude that the radiative proton capture on $^{22}$Mg can not account for the depletion of $^{22}$Na in classical novae.\newline
\indent It was known from previous work \cite{trac01,trac06} that one-proton nuclear breakup reactions of rare isotope beams can provide important information needed to determine the astrophysical reaction rates for radiative proton capture reactions that are outside the reach of other direct or indirect methods, by extracting parameter-free the asymptotic normalization coefficients. This information replaces and/or complements the information obtained from transfer reactions (the ANC method \cite{azha01}) that would require radioactive beams at lower energies of much better purity and intensity.  We demonstrate with this work the extension of the method from light $p-shell$ to mid $sd-shell$ nuclei, and the advantage that it can be used for beams of lower quality, such as cocktail beams with intensities as low as $\sim$ 100 particles per second. However, as one goes higher in mass, configuration mixing may play a more important role and one must disentangle various configurations, as we did for $^{23}$Al, using exclusive measurements involving core-$\gamma$-ray coincidences. Only after that, the data can be used to extract astrophysical $S$ factors and evaluate reaction rates for radiative proton captures. As shown here, in addition to good quality experimental data, reliable cross section calculations are necessary.\newline

\begin{acknowledgments}
\indent We thank the GANIL operating staff for the excellent beam conditions. The work was supported in part by the US DOE under Grant 
DE-FG02-93ER40773, the Robert A. Welch Foundation under Grant A-1082 and the EURONS program. M. H. acknowledges the NSF grant PHY-0758099. F. C. acknowledges the CNCSIS (Romania) grant PN-II-PCE-2007-1/258. A. Banu would like to thank C. A. Gagliardi for useful discussions and his interest in this work. 
\end{acknowledgments}

\bibliography{Al23_paper_refs}% Produces the bibliography via BibTeX.

\end{document}